\documentclass[a4paper]{jpconf}
\usepackage{graphicx}
\usepackage{amssymb}  
\usepackage{amsmath}  
\usepackage{latexsym} 
\usepackage{amsthm}   
\begin{document}
\title{Density-dependent nucleon-nucleon interaction from UIX three-nucleon force}
\author{Alessandro Lovato$^{1}$,Omar Benhar$^{2,3}$,Stefano Fantoni$^{1,4}$,Alexey Yu. Illarionov$^{5}$ and Kevin E Schmidt$^{6}$}
\address{$^1$ SISSA and INFN, Sezione di Trieste. I-34014 Trieste, Italy}
\address{$^2$ INFN, Sezione di Roma. I-00185 Roma, Italy}
\address{$^3$ Dipartimento di Fisica, Universit\`a ``La Sapienza''. I-00185 Roma, Italy}
\address{$^4$ CNR-DEMOCRITOS National Supercomputing Center. I-34014 Trieste, Italy}
\address{$^5$ Dipartimento di Fisica, Universit\`a di Trento. I-38123 Povo, Trento, Italy}
\address{$^6$ Department of Physics, Arizona State University, Tempe, AZ 85287}
\ead{lovato@sissa.it}
\begin{abstract}
A density-dependent two-nucleon potential has been derived in the formalism of correlated basis function. The effects of $3$-particle interactions has been included by integrating out the degrees of freedom of the third nucleon. The potential can be easily employed in nuclear matter calculations. It yields results in agreement with those obtained from the underlying three-body potential. The use of the density dependent potential allowed us to study the effects of three-nucleon interactions in symmetric nuclear matter within the Auxiliary Field Diffusion Monte Carlo (AFDMC) computational scheme.
\end{abstract}
\section{Introduction}

The Auxiliary Field Diffusion Monte Carlo (AFDMC) \cite{Schmidt1999} results of Ref. \cite{AFDMC1} do not lead to a decrease of the binding energy predicted by Fermi-Hyper-Netted-Chain (FHNC) and Brueckner-Hartree-Fock (BHF) calculations \cite{FHNC-BHF}. This suggested that the overstimation of the SNM binding energy resulting from FHNC calculation with UIX three body potential \cite{UIX}, may not be solved with a better choice of the variational wave function.   

More sofisticated three body potentials, while providing a quantitative account of the energies of the ground and low-lying excited  states of nuclei with $A\leq12$ \cite{steve_RNC}, give very different results in pure neutron matter (PNM) \cite{Sarsa2003}. 

Lagaris and Pandharipande \cite{LP1} and Friedman and Pandharipande \cite{FrP} constructed an effective, density-dependent, two-nucleon potential (TNI) to take into account three- and many-nucleon forces. However, they adopted a purely phenomenological procedure, as the parameters of the density dependent functions appearing in this potential were determined through a fit of the saturation density, the binding energy per nucleon and the compressibility of symmetric nuclear matter (SNM). In general, fitting the potential to results obtained with many-body techniques makes the potential itself affected by the approximations of the many body calculations.

Our potential \cite{lovato_11} improves the TNI model since it has been derived from a realistic microscopic three-nucleon force, the UIX potential, which provides a good description of the properties of light nuclei, although very recently it has been shown \cite{kievsky} that UIX does not reproduce both the $nd$ scattering lenght and the vector polarization observables $A_y$ and $iT_{11}$. However, our method can be applied to more refined three nucleon interactions, like the chiral NNLOL described in \cite{kievsky}.

To obtain the density dependent potential, the average on the degrees of freedom of the third particle has been carried out using a formalism suitable to account for the full complexity of nuclear dynamics. Our results show that, in doing such reduction, of great importance is the proper inclusion of both dynamical and statistical NN correlations, whose effects on many nuclear observables have been found to be large \cite{RMP1,Pand1997}. 

We have used CBF and the Fantoni-Rosati (FR) cluster expansion  formalism\cite{FR_cluster} to perform the calculation of the linear-in-density terms of the effective potential, arising from the irreducible three-nucleon interactions modeled by the UIX potential.

The effective potential has been implemented in the AFDMC computational scheme to obtain the EoS of SNM, while similar calculations using the UIX potential are not yet possible, due to the complexities arising from the commutator term. In addition, the density-dependent potential can be used to include the effects of three-nucleon interactions in the calculation of the nucleon-nucleon scattering cross section in the nuclear medium. The knowledge of this quantity is required to obtain a number of nuclear matter properties of astrophysical interest, ranging from the transport coefficients to the neutrino emission rates \cite{BV,BFFV}.

\section{Two- and three- nucleon forces}
\label{TBF}
The Argonne $v_{8}$ \cite{V8P} two-body potential model is given by 
\begin{equation}
\hat{v}_{ij}=\sum_{p=1}^8 v^p(r_{ij})O^{p}_{ij}\, 
\label{eq:av8_potential}
\end{equation}
where
\begin{equation}
O^{p=1-8}_{ij}=(1,\sigma_{ij},S_{ij},\mathbf{L}_{ij}\cdot\mathbf{S}_{ij})\otimes(1,\tau_{ij})\,.
\end{equation}
In the above equation,  $\sigma_{ij}={\boldsymbol \sigma}_i \cdot {\boldsymbol \sigma}_j$ and 
$\tau_{ij}={\boldsymbol \tau}_i \cdot {\boldsymbol \tau}_j$, where ${\boldsymbol \sigma}_i$ and ${\boldsymbol \tau}_i$ are Pauli matrices acting on the spin or isospin of the $i$-th particle, while $S_{ij}$ is the tensor operator, $\mathbf{L}_{ij}$ is the relative angular momentum 
and $\mathbf{S}_{ij}$ is the total spin of the pair.

We used the so called Argonne $v_{8}^\prime$ and Argonne $v_{6}^\prime$ potentials, which are not simple truncations of the Argonne $v_{18}$ potential \cite{av18}, but rather reprojections \cite{V6P} obtained by refitting the scattering data and the deuteron binding energies.
In all light nuclei and nuclear matter calculations the results obtained with the $v_{8}^\prime$ are very close to those obtained with the full $v_{18}$.

Using a nuclear Hamiltonian including only two-nucleon interactions leads to the underbinding of light nuclei and overestimating the equilibrium density of nuclear matter. Hence, the contribution of three-nucleon interactions must be taken into account, by adding to the Hamiltonian  the corresponding potential. One of the most widely used is Urbana IX (UIX) \cite{UIX} potential, consisting of two terms: the  Fujita and Miyazawa \cite{Fujita_Miyazawa} attractive two-pion exchange interaction $V^{2\pi}$ and the purely phenomenological repulsive term $V^R$
\begin{align}
\hat{V}^{2\pi}&=A^{2\pi}\sum_{cyclic}\Big(\{\hat{X}_{ij},\hat{X}_{jk}\}\{\tau_{ij},\tau_{jk}\}
+\frac{1}{4}[\hat{X}_{ij},\hat{X}_{jk}][\tau_{ij},\tau_{jk}]\Big)\nonumber \\
V^R&=U_0\sum_{cyclic}T^2(m_\pi r_{ij})T^2(m_\pi r_{jk})\, .
\end{align}
The spin structure of $V^{2\pi}$ is given by
\begin{equation}
\hat{X}_{ij}=Y(m_\pi r)\sigma_{ij}+T(m_\pi r)S_{ij}\, .
\end{equation}
The radial functions read
\begin{align}
Y(x)=\frac{e^{-x}}{x}\xi_Y(x)\quad,\quad T(x)=\Big(1+\frac{3}{x}+\frac{3}{x^2}\Big)Y(x)\xi_T(x)\, 
\label{eq:YT}
\end{align}
where $\xi_{Y}(x)=\xi_{T}(x)=1-\exp(-cx^2)\,$ are short-range cutoff functions. The parameters $A_{2\pi}$ and $U_0$ are varied to fit the observed binding energies of $^3$H and $^4$He and to reproduce the empirical nuclear matter saturation density, while the cutoff parameter is kept fixed at $c=2.1$ fm$^{-2}$.

\section{Correlated Basis Theory and Cluster Expansion technique}
In the correlated basis theories of Fermi liquids \cite{clark,OCBfantoni}, the expectation 
value of the two-body potential can be written in the form
\begin{eqnarray}
\langle \hat{v} \rangle =\frac{1}{2}\rho\sum_p \int d\vec{r}_1 d\vec{r}_2 v^{\,p}_{12}\,g^{\, p}_{12} \, ,
\label{eq:two_body_exp}
\end{eqnarray}
where
\begin{equation}
g_{12}^p=\frac{A(A-1)}{\rho^2}\frac{\text{Tr}_{12}\int dx_3 \ldots  dx_A \Phi_0^*F^\dagger O_{12}^p F \Phi_0}{\int dX  \ \Phi_0^*F^\dagger F \Phi_0}\, ,
\label{eq:g2_def}
\end{equation}
are the operatorial components of the two--body distribution function. The uncorrelated wavefunction $\Phi_0$ in nuclear matter is conveniently chosen to be a Slater determinant of plane waves.

The structure of the correlation operator $\hat{F}$ reflects the complexity of the Argonne $v_{6}^\prime$ nucleon-nucleon potential \cite{wiringa_pandha_1}:
\begin{equation}
F=\mathcal{S} \prod_{j>i=1}^A F_{ij} \qquad \text{with} \qquad \hat{F}_{ij} = \sum_{p=1}^6 f^{p}(r_{ij})\hat{O}^{p}_{ij} \, .
\label{eq:Foperator}
\end{equation}

The radial functions $f^{p}(r_{ij})$, appearing in the definition of the correlation operator are determined by the minimization of the energy expectation value $E_V=\langle\Psi_0|H|\Psi_0\rangle$, which provides an upper bound to the true ground state energy $E_0$. As explained in detail in \cite{lovato_11}, the calculation of $E_V$ in CBF theories is carried out by i) expanding $\langle\Psi_0|H|\Psi_0\rangle$ in powers of 
dynamical correlations $h(r_{ij})=f^{c}(r_{ij})^2-1$, $2f^{c}(r_{ij})f^{p}(r_{ij})$, $f^{p>1}(r_{ij})f^{q>1}(r_{ij})$ that vanish in uncorrelated matter and ii) summing up the main series of the resulting cluster terms by solving a set of coupled integral equations. To accomplish the first of these two steps, an extension of the FR cluster expansion \cite{FR_cluster}, that is able to treat scalar correlations only, was introduced in \cite{wiringa_pandha_1} to deal with spin--isospin dependent correlation operators, like those of Eq. (\ref{eq:Foperator}). 

The cluster terms are most conveniently represented by diagrams \cite{wiringa_pandha_1}. 
consisting of dots (vertices) connected by different kinds of correlation lines. In particular $h_{ij}$ is usually represented by a dashed line, $2f^{c}_{ij}f^{p}_{ij}$ by a single wavy line, and $f^{p>1}_{ij}f^{q>1}_{ij}$ by a doubly wavy line. In addition to the dynamical correlation lines, 
there are also statistical correlation lines, represented by solid oriented lines forming closed loops that never touch each other. They arise from the Slater determinant and are associated with the Slater function $\ell_{ij}$
\begin{equation}
\ell(k_Fr_{ij}) = 3\Big[\frac{\sin(k_Fr_{ij})-k_Fr_{ij}\cos(k_Fr_{ij})}{(k_Fr_{ij})^3}\Big]\, .
\end{equation}
Open dots of the diagrams represent the active (or interacting) particles (1 and 2), while black dots are associated with passive particles, i.e. those in the medium. Integration over the coordinates of the passive particles leads to the appearance of a factor $\rho$.

It can be shown that all the diagrams contributing to $E_V$, hence to $\langle \hat{v} \rangle$, are linked. Those built with scalar passive bonds only,  with the only exception of the so called {\sl elementary} diagrams, can be summed up in closed form by solving the FHNC equations \cite{FR_cluster}.

On the other hand, diagrams having one or more passive operatorial bonds are calculated at leading order only. This implies that at most two operatorial passive bonds can be attached to any internal point, thus only diagrams with Single Operator Chain (SOC) are considered. Such an approximation is justified by the observation that operatorial correlations are much weaker than the scalar ones.

\section{Density dependent potential}
As for the two-body potential $v_{8}$, it is possible to write the UIX three body potential in the following way 
 \begin{equation}
\label{expand:v123}
V_{123} \equiv \sum_p V_{123}^p {O}_{123}^p\, \, .
\end{equation}
The expectation value of $\hat{V}_{123}$ then reads
 \begin{equation}
\frac{\langle V\rangle}{A}=\frac{1}{3!}\ \rho^2 \sum_P \int dr_{12}dr_{13} V_{123}^p\,g_{123}^p \, ,
\label{eq:3_body_exp_with_g3}
\end{equation}
with
\begin{equation}
g_{123}^p=\frac{A!}{(A-3)!}\frac{\text{Tr}_{123}\int dx_4 \ldots dx_A \Phi^\dagger_0F^\dagger {O}_{123}^p 
F \Phi_0}{ \rho^3\int dX \  \Phi^\dagger_0F^\dagger F \Phi_0}\, .
\label{eq:g3_def}
\end{equation}

\begin{figure}[!t]
\begin{center}
\includegraphics[angle=0,width=8.0cm]{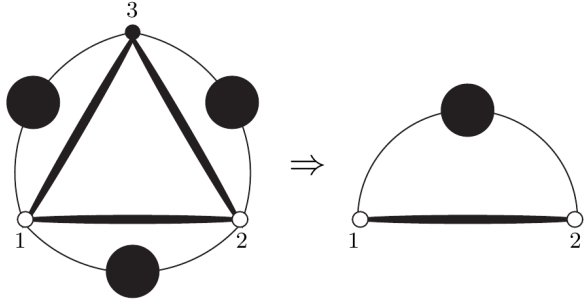}
\caption{Diagrammatic representation of Eq. (\ref{eq:ddp_request}). }
\label{fig:g3_g2}
\end{center}
\end{figure}

We require that the expectation values of $V_{123}$ and of $v_{12}(\rho)$ be the same, implying in turn
\begin{equation}
\sum_{P}\frac{\rho}{3}\int d\vec{r}_3 V_{123}^P\,g_{123}^P=\sum_p v_{12}^{p}(\rho)\,g_{12}^p\, .
\label{eq:ddp_request}
\end{equation}
A diagrammatic representation of the above equation, which should be regarded as the definition of  the $v_{12}(\rho)$, 
is shown in Fig. \ref{fig:g3_g2}. The graph on the left-hand side represents the three-body potential times the three-body correlation function, integrated over the coordinates of particle $3$. Correlation and exchange lines are schematically depicted with a line having a bubble in the middle, while the thick solid lines represent the three-body potential. The diagram in the right-hand side 
represents the density-dependent two-body potential, dressed with the two-body distribution function. 
Obviously, $v_{12}^\rho$ has to include not only the three-body potential, but also the effects of correlation and 
exchange lines.  

\begin{figure}[!t]
\begin{center}
\includegraphics[angle=0,width=12.0cm]{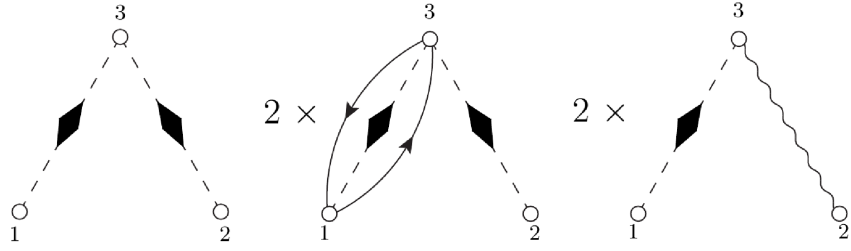}
\caption{Diagrams contributing to the density-dependent potential. The dashed lines with diamonds represent the first order approximation to $g_{bose}^{NLO}(r_{ij})$, discussed in the text.  \label{fig:two_body_bose_corr}}
\label{fig:relevant_diagrams}
\end{center}
\end{figure}

In the construction of the density dependent interaction we have found that the most relevant diagrams are those depicted in Fig. \ref{fig:relevant_diagrams} that indeed include both statistical and dynamical correlations. To simplify the pictures, the three-body potential  acting on particles $1$, $2$ and $3$ is not explicitely depicted. In order to include higher order cluster terms, we have replaced the scalar correlation line ${f^c_{ij}}^2$ with the Next to Leading Order (NLO) approximation to the bosonic two-body correlation function:
\begin{equation}
{f^{c}_{ij}}^2\rightarrow g_{bose}^{NLO}(r_{ij})={f^{c}_{ij}}^2\Big(1+\rho\int d\vec{r}_3 h_{13} h_{23}\Big)\, .
\label{eq:g12_NLO}
\end{equation}
\begin{figure}[!b]
\begin{center}
\includegraphics[angle=270,width=7.5cm]{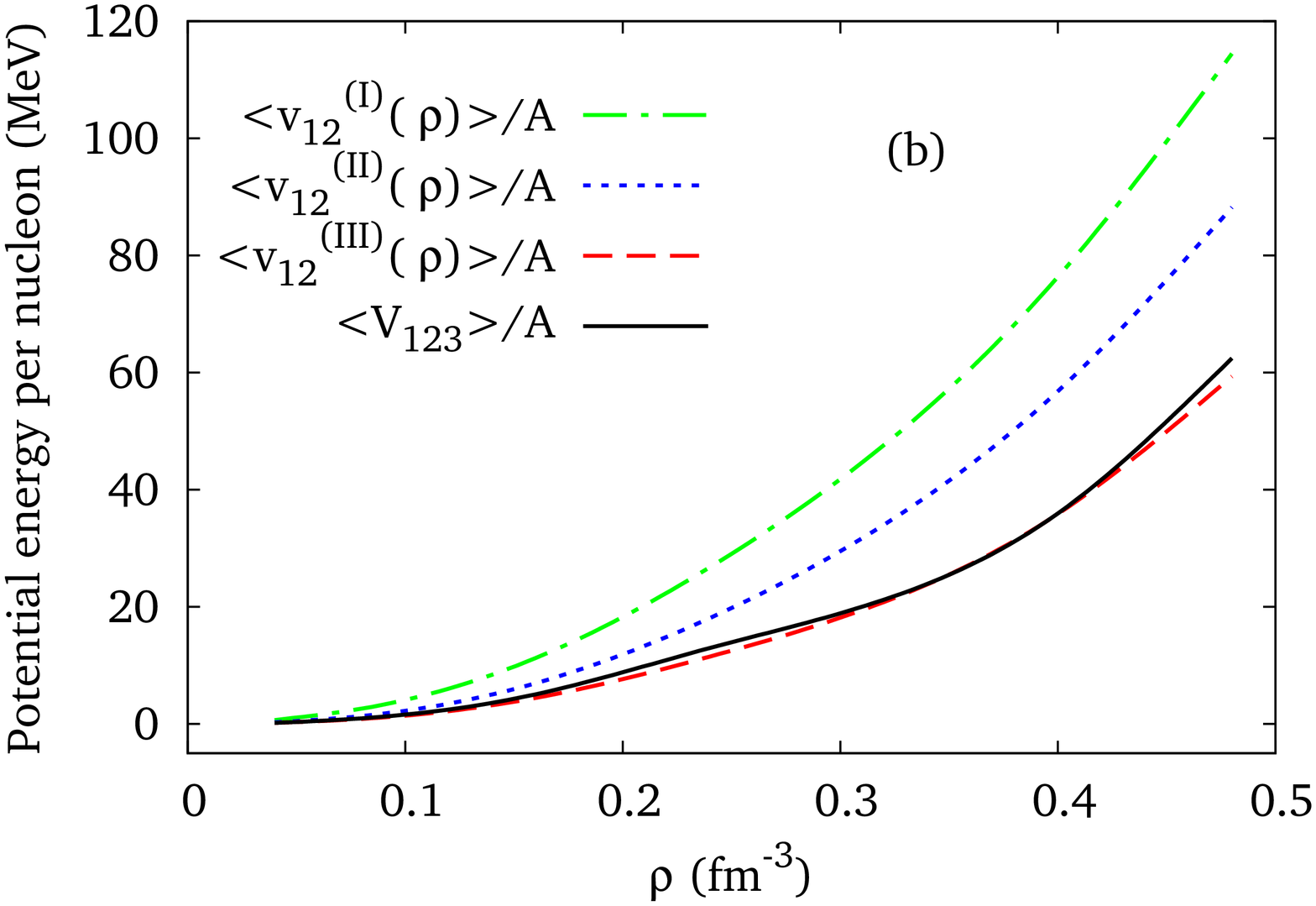}
\includegraphics[angle=270,width=7.5cm]{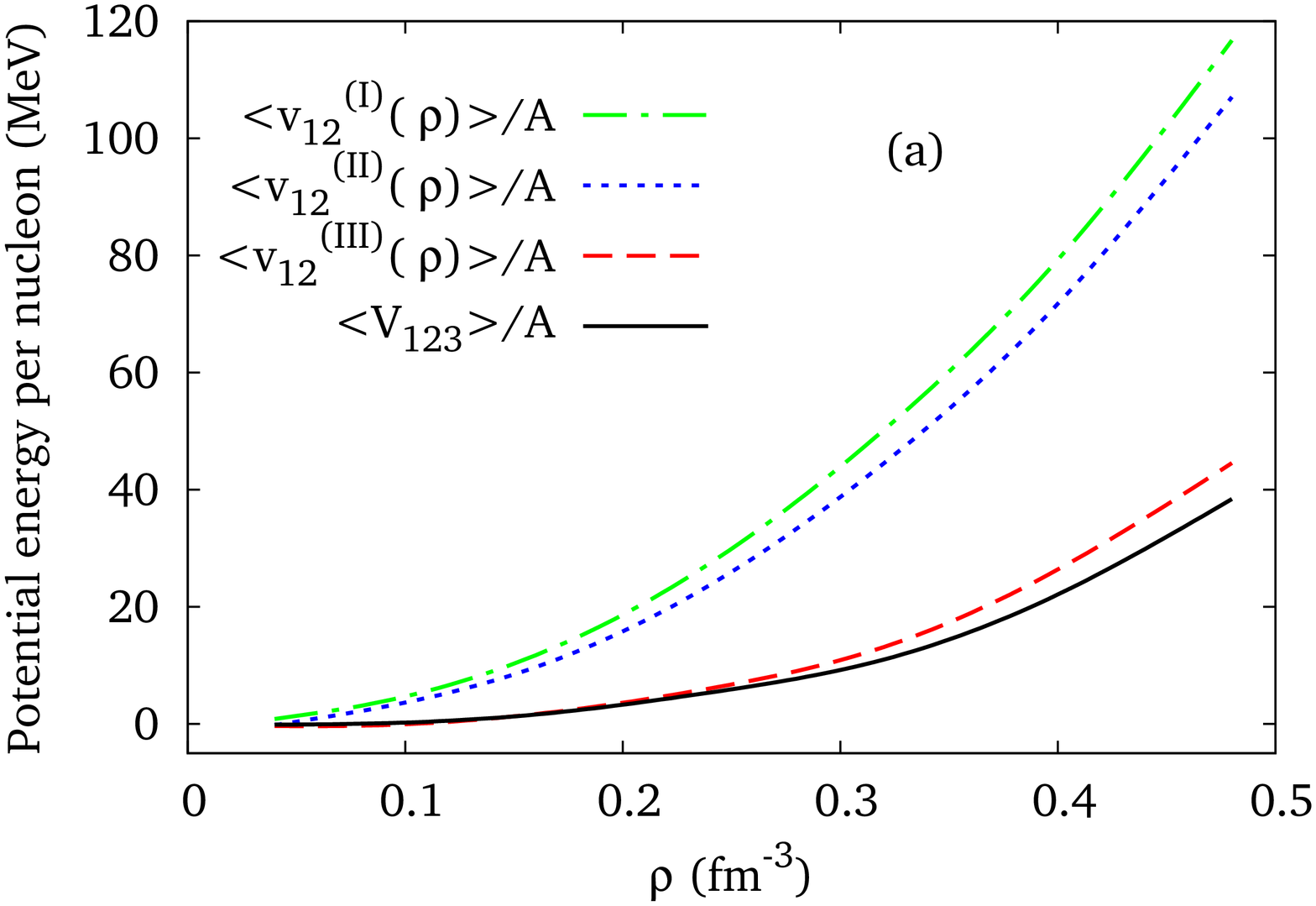}
\caption{Contributions of the density-dependent potential to the energy per particle of SNM (a) and PNM (b), compared to 
the expectation value of the three-body potential UIX: $\langle V_{123}\rangle/A$. \label{fig:compare_potentials}}
\end{center}
\end{figure}
The inclusion of both statistical and dynamical correlations plays a fondamental role in the determination of a realistic density dependent interaction, as can be seen from Fig \ref{fig:compare_potentials}, where different density dependent potentials contributions are plotted and compared with the full calculation with the genuine three-nucleon interaction. In particular $v_{12}^I(\rho)$, obtained by averaging over the third particle without statistical and dynamical correlations is far from the solid curve of UIX. When only statistical correlations are considered, the relative potential, $v_{12}^{II}(\rho)$, comes closer to the UIX curve. Only when the full set of diagrams of Fig. \ref{fig:relevant_diagrams} is included in the calculation
\begin{align}
v^{(III)}_{12}(\rho)&=\frac{\rho}{3}\int d x_3\,V_{123}\,\Big[g_{bose}^{NLO}(r_{13}) g_{bose}^{NLO}(r_{23})
(1-2P_{13}\ell_{13}^2)+4g_{bose}^{NLO}(r_{13})f_{c}(r_{23})\hat{f}(r_{23})\Big]\, ,
\label{eq:ddp_diagrams_expression}
\end{align}
where $\hat{f}(r_{23})$ denotes the sum of non central correlations, the contribution of the density dependent potential comes very close to that of the original UIX.
A proper treatment of the statistical correlation is crucial for the construction of a density dependent potential to be treated as an additive term to the standard two-body potential. Some care is needed for the density dependent potential to reproduce the exchange loop involving particles $1$, $2$ and $3$ with the appropriate symmetry factor (see \cite{lovato_11} for a thorough discussion).

\section{Numerical results: PNM and SNM equation of state}
As explained in \cite{lovato_11}, the correlation functions are determined by minimizing the two-body cluster term of the energy expectation value for a set of parameters $d_c$, $d_t$, $\beta_p$ and $\alpha_p$. The energy expectation value $E_V$ calculated in full FHNC/SOC approximation has then to be minimized with respect to the variation of these parameters. To this aim a “Simulated annealing” \cite{SA} optimization algorithm has been implemented \cite{lovato_11}. 

Both SOC approximation and the fact that elementary diagrams are neglected may lead to a violation of the variational principle. To keep this effect under control, an additional constraint on the kinetic energy with respect to the optimization performed in \cite{fabrocini_wiringa} has been considered.

\begin{figure}[!h]
\begin{center}
\includegraphics[angle=270,width=7.5cm]{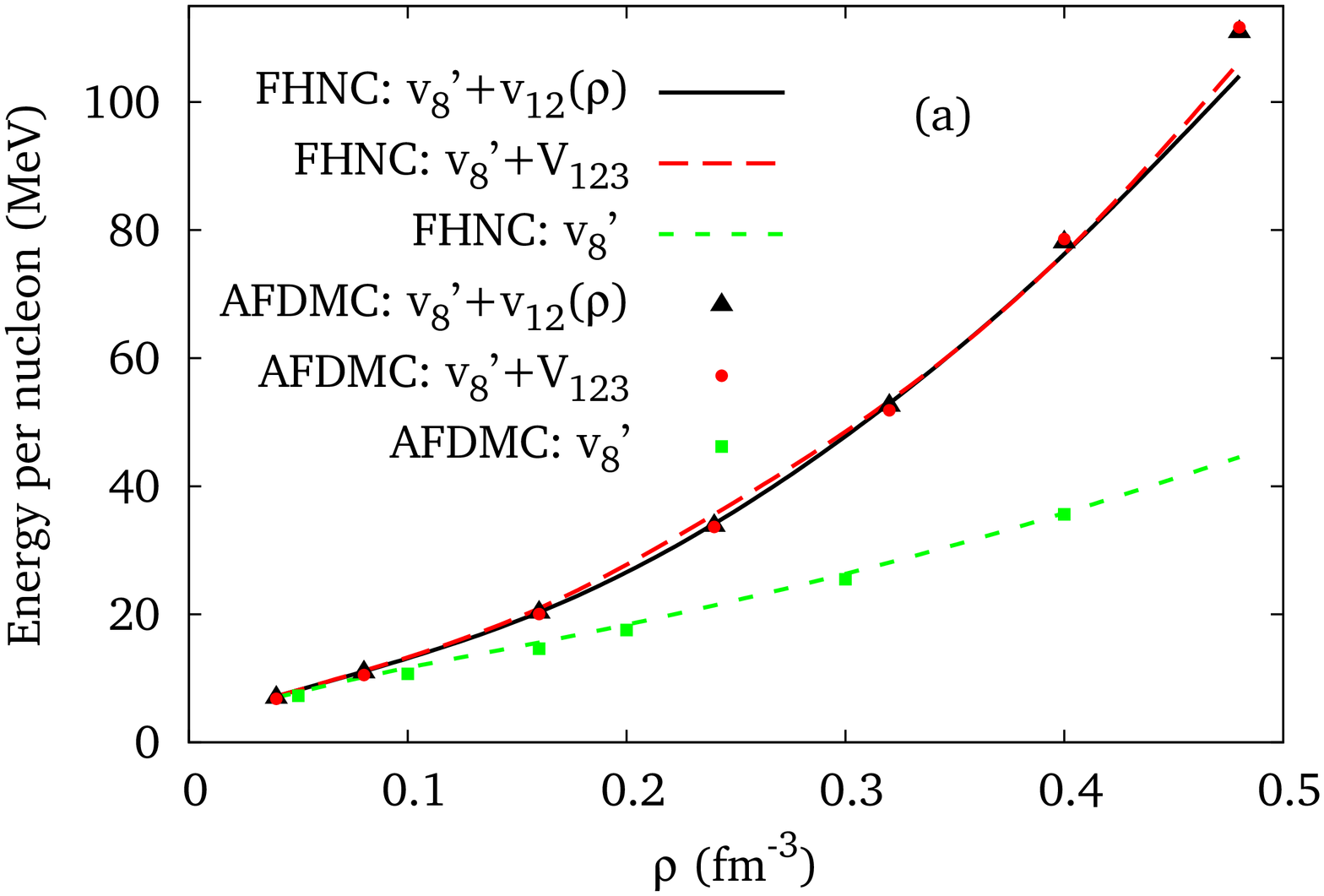}
\includegraphics[angle=270,width=7.5cm]{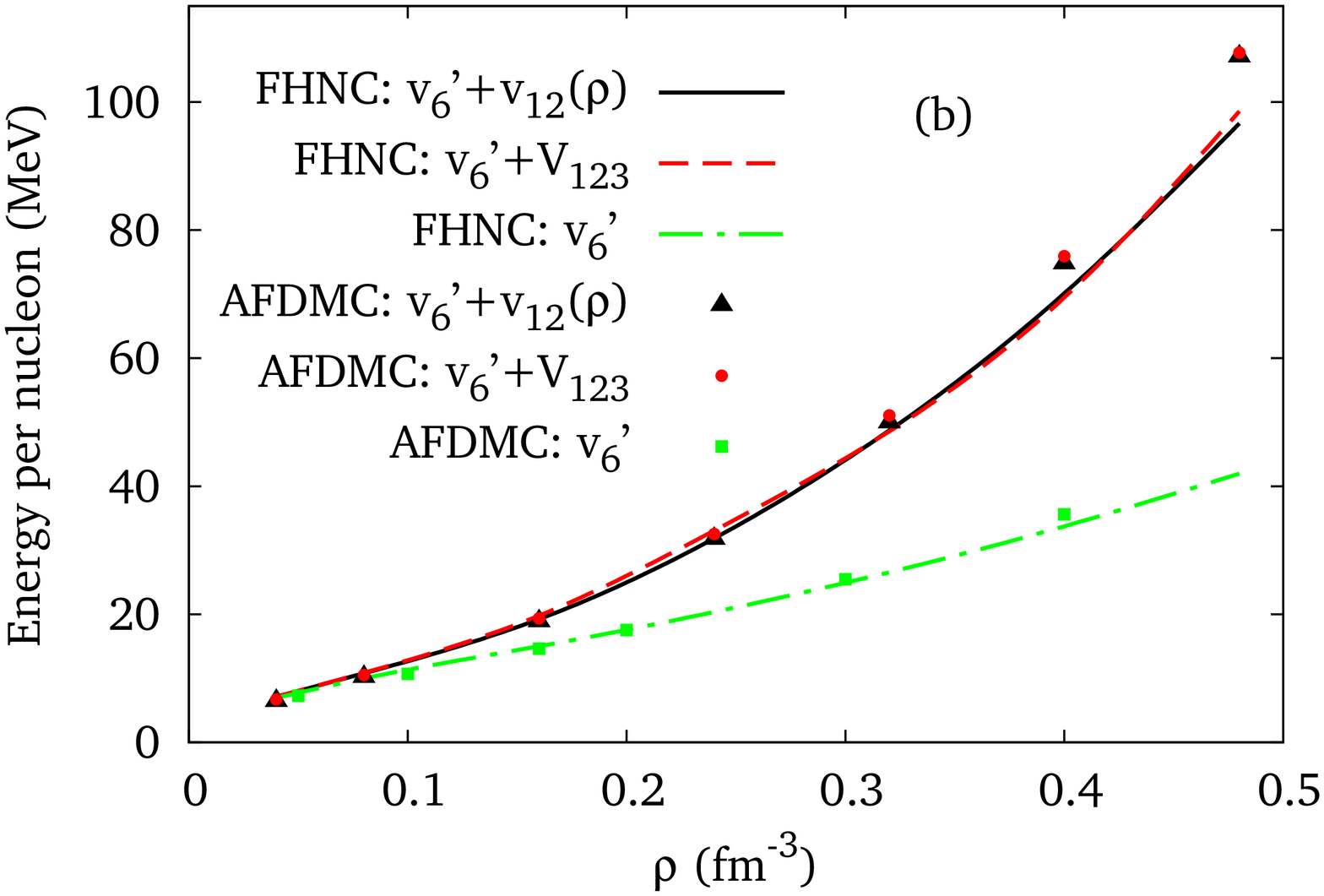}
\caption{Energy per particle for PNM, obtained using the density-dependent potential added to the Argonne $v_{8}^\prime$ (a) and to Argonne $v_{6}^\prime$ (b) potentials. The energies are compared to those obtained from the genuine three-body potential and from 
the two-body potentials alone.  \label{fig:eos_pnm}}
\end{center}
\end{figure}

We carried out AFDMC simulations \cite{Schmidt1999} for PNM with $A=66$ and SNM with $A=28$ nucleons in a periodic box. The finite-size errors in PNM simulations have been investigated in \cite{Gandolfi2009} by comparing the Twist Averaged Boundary Conditions (TABC) with the Periodic Box Condition (PBC). The kinetic energy of 66 fermions approaches the thermodynamic limit very well, so that the energies of 66 neutrons computed using either TABC or PBC turn out to be almost the same. We can infer that the finite-size errors in the present AFDMC calculations for PNM do not exceed 2\% of the asymptotic value of the energy.

The finite-size effects of SNM calculations can be estimated from the difference of the energies of PNM obtained with 14 neutrons and the TABC asymptotic value, which is of the order of 7\%.

\begin{figure}[!ht]
\begin{center}
\includegraphics[angle=270,width=7.5cm]{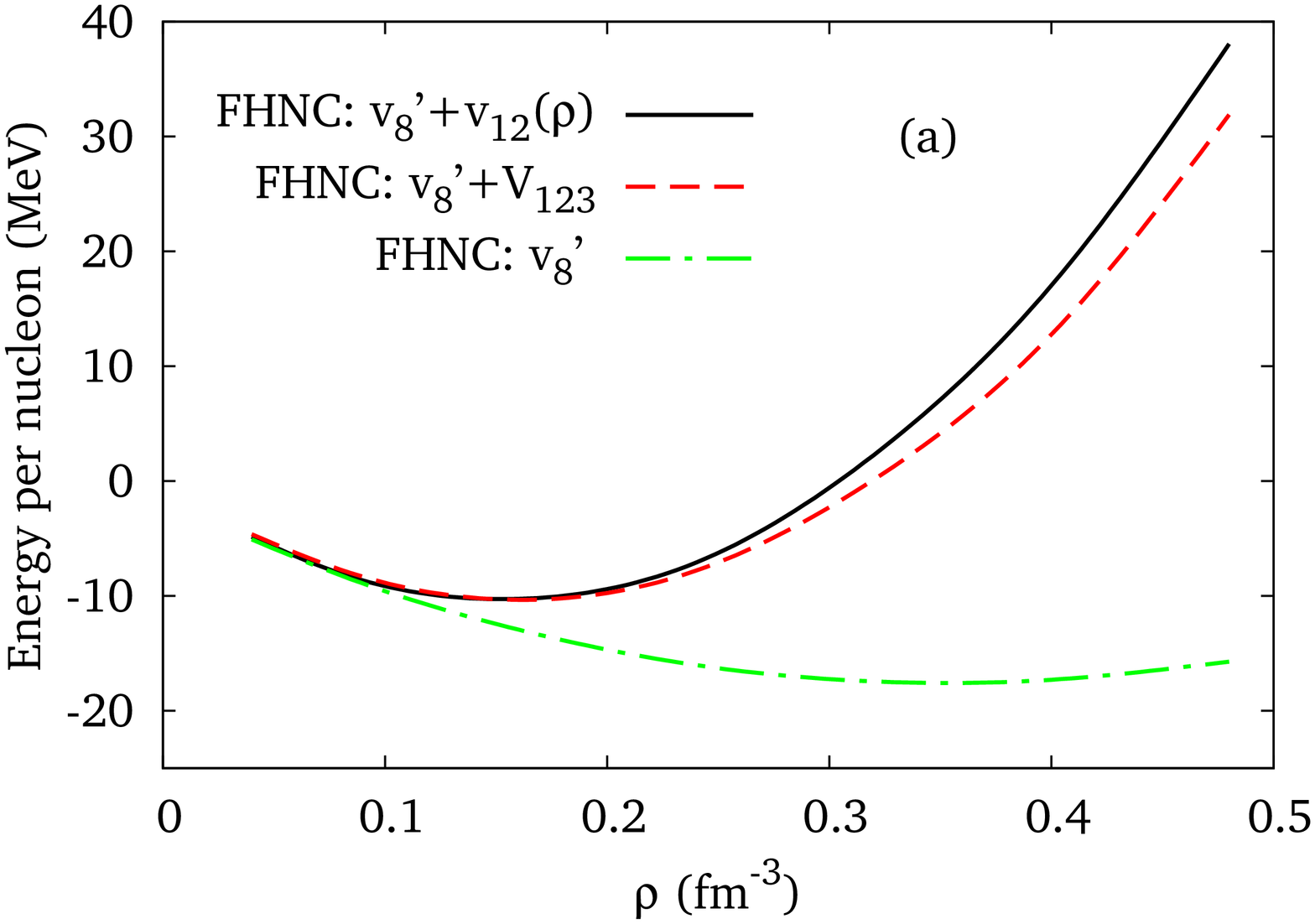}
\includegraphics[angle=270,width=7.5cm]{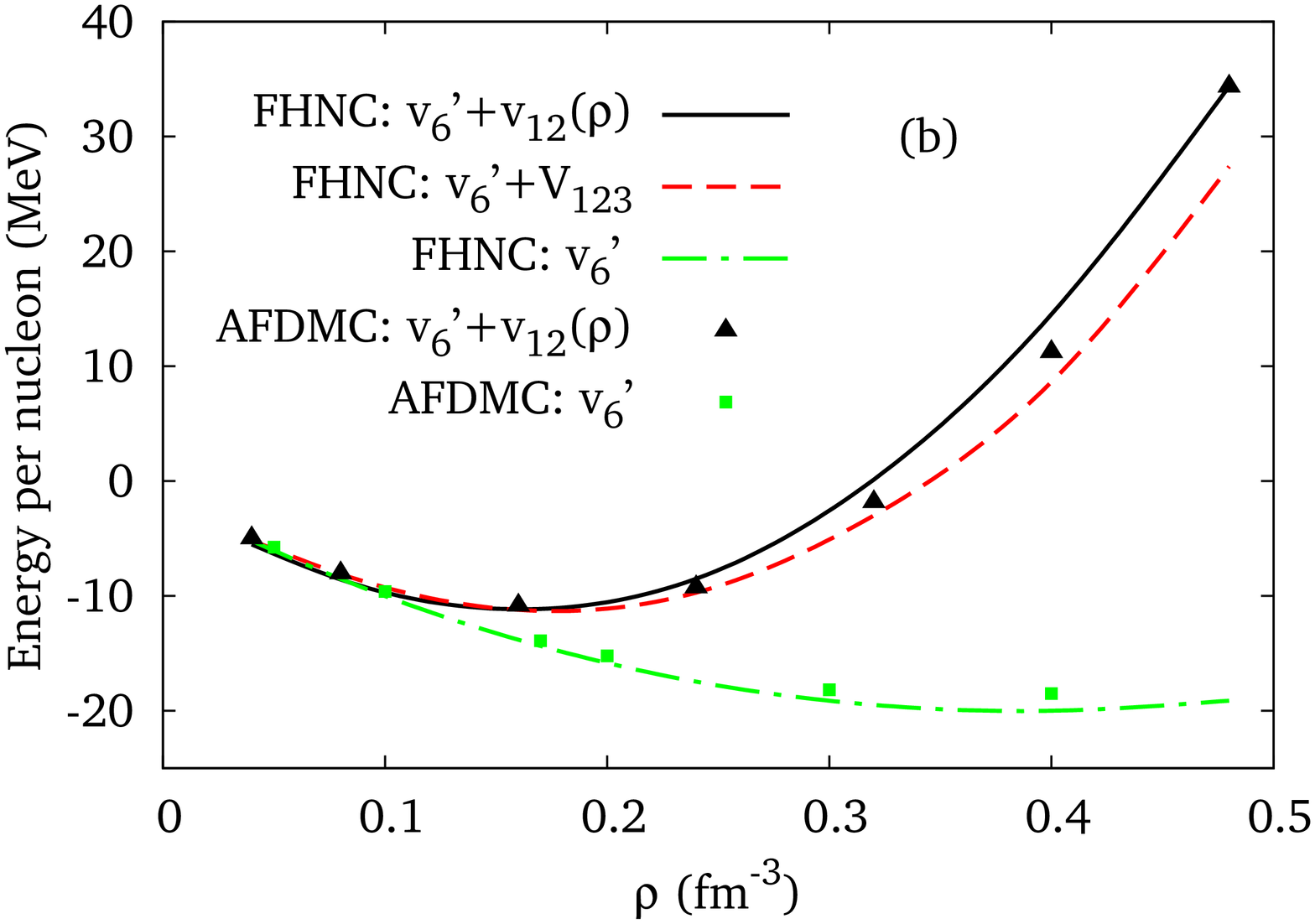}
\caption{Same as in Fig. \ref{fig:eos_pnm}, but for SNM
\label{fig:eos_snm}}
\end{center}
\end{figure}

FHNC/SOC and Monte Carlo calculations provide very close results, as shown in Figs. \ref{fig:eos_pnm} and \ref{fig:eos_snm}, to be compared with those of Ref. \cite{AFDMC1} where the agreement between FHNC and Monte Carlo methods were not nearly as good.

The EoS of PNM, displayed in Fig. \ref{fig:eos_pnm}, obtained with the three-body potential UIX and using the density-dependent two-body potential are very close to each other. In the same figure, for the sake of comparison, we also report the results of calculations carried out including the two--body potential only.

Despite cluster contributions proportional to $\rho^2$ have been neglected in the density dependent potential, with the exception of the line with diamonds of Fig. \ref{fig:two_body_bose_corr}, even at high densities the EoS relative to $v_{12}(\rho)$ remain close to those obtained with UIX. Probably, in this case a compensation among second and higher order terms takes place.

The density-dependent potential has been also employed in AFDMC calculations. As can be seen in Fig. \ref{fig:eos_pnm}, the triangles representing the results of  this calculation are very close, when not superimposed, to the circles corresponding to the UIX three-body potential AFDMC results.

For what concern the EoS of symmetric nuclear matter, see Fig. \ref{fig:eos_snm}, at densities lower than $\rho=0.32\,\text{fm}^{-3}$, the curves resulting from UIX and the density-dependent potential are very close to one other, while for $\rho>0.32\,\text{fm}^{-3}$ a gap between them appears. 

\begin{table}[!h]
\begin{center}
\caption{Values for the saturation densities, the binding energy per particle, and the compressibility of SNM relative to the EoS of Fig. \ref{fig:eos_snm}. \label{table:parameters_eos}}
\vspace{0.3cm}
\begin{tabular}[c]{ c c c c c } 
\hline 
FHNC/SOC & $v_{6}^\prime + V_{123}\quad$ & $ v_{6}^\prime + v(\rho) $ & $ v_{8}^\prime + V_{123} $ & $ v_{8}^\prime + v(\rho)$\\ 
\hline
$\rho_0$ (fm$^{-3}$)& 0.17 & 0.16 & 0.16 & 0.15 \\ 

$E_0$ (MeV) &-11.3 & -11.2 & -10.3 & -10.3 \\ 

K (MeV) & 205 & 192 & 189 & 198 \\ 
\hline
\end{tabular} 
\hspace{0.5cm}
\begin{tabular}[c]{c c} 
\hline 
AFDMC & $ v_{6}^\prime + v(\rho)$ \\ 
\hline
$\rho_0$ (fm$^{-3}$)& 0.17  \\ 
$E_0$ (MeV) &-10.9\\ 
K (MeV) & 201 \\ 
\hline
\end{tabular} 
\end{center}
\end{table}
In Table \ref{table:parameters_eos}, the saturation density $\rho_0$, the binding energy per particle $E(\rho_0)$ and the compressibility $K = 9\rho_0 (\partial E(\rho)/\partial\rho)^2$ for all the EoS of Fig. \ref{fig:eos_snm} are listed. It is remarkable that the values obtained with density dependent potential are very close to those resulting from the genuine UIX three body potential. While the saturation density is well reproduced, this being not surprising since the constant $U_0$ of the UIX potential has been fitted to this value, the values of the compressibility
are slightly lower than the experimental ones. AFDMC calculations lower the variational results for $E_0$ by $0.6\,\text{MeV}$ only, showing that the binding energy overstimation of the FHNC/SOC calculations performed with UIX potential can not be solved with a better choice of the variational wavefunction.

\section{Conclusions}
We have developed a novel scheme, suitable to obtain an effective density-dependent NN potential taking into account the effects of three-nucleon interactions. Our approach is fully consistent with the treatment of correlations underlying the FHNC and AFDMC approaches.

The PNM and SNM equation of state resulting from the density-dependent potential turn out to be very close to those obtained with the UIX three-body potential. In this context, a critical role is played by the treatment of both dynamical and statistical correlations, a distinctive feature of our approach, as compared to different reduction schemes based on effective interactions \cite{hebeler_schwenk,weise}.

For the first time, a AFDMC calculation of the equation of state of SNM consistently including the effects of three nucleon forces has been carried out. The results of this calculation show that the $v_6^\prime + $UIX hamiltonian, or equivalently the one including the effective potential, fails to reproduce the empirical data. The discrepancy has most likely to be ascribed either to deficiencies of the UIX model or to the effect of interactions involving more than three nucleons. 

As a further development, we are studying the dependence on the specific model of three-nucleon force \cite{lovato_11b}, as well as the inclusion of four- and many-nucleon interactions, whose effects are expected to be critical for the determination of the properties of high density neutron star matter .

Our density dependent potential could be easily employed in many-body approaches other than those based on the CBF formalism or quantum  Monte Carlo simulations, such as the G-matrix and self-consistent Green function theories \cite{Baldo, Polls, Dickhoff}.

\end{document}